\documentclass{article}
\usepackage{spconf}
\usepackage{amsmath,graphicx}
\usepackage{booktabs}
\usepackage{cite}
\usepackage{multirow}
\usepackage{multicol}
\usepackage{tabularx}
\usepackage{enumitem}
\usepackage{amsfonts,amssymb}
\usepackage{cleveref}
\usepackage{amsfonts}
\usepackage{upgreek}
\usepackage{float}
\usepackage{subcaption}
\usepackage{graphicx}
\usepackage{makecell}
\usepackage{array}
\usepackage{units}
\usepackage[square,numbers]{natbib}
\usepackage{caption, setspace}
\usepackage[activate={true,nocompatibility},final,tracking=true,kerning=true,spacing=true,factor=1100,stretch=10,shrink=10]{microtype}
\usepackage[hidelinks]{hyperref}

\title{An Approach to Super-Resolution of Sentinel-2 Images Based on Generative Adversarial Networks}
\name{Kexin Zhang~$^{1}$, Gencer Sumbul~$^{2}$, Beg\"{u}m Demir~$^{2}$}
\address{$^{1}$~Shanghai Jiao Tong University, $^{2}$~Technische Universit\"at Berlin
}

\begin{document}
%\ninept
%
\maketitle
\begin{abstract}
This paper presents a generative adversarial network based super-resolution (SR) approach (which is called as S2GAN) to enhance the spatial resolution of Sentinel-2 spectral bands. The proposed approach consists of two main steps. The first step aims to 
increase the spatial resolution of the bands with 20m and 60m spatial resolutions by the scaling factors of 2 and 6, respectively. To this end, we introduce a generator network that performs SR on the lower resolution bands with the guidance of the bands associated to 10m spatial resolution by utilizing the convolutional layers with residual connections and a long skip-connection between inputs and outputs. The second step aims to distinguish SR bands from their ground truth bands. This is achieved by the proposed discriminator network, which alternately characterizes the high level features of the two sets of bands and applying binary classification on the extracted features. Then, we formulate the adversarial learning of the generator and discriminator networks as a min-max game. In this learning procedure, the generator aims to produce realistic SR bands as much as possible so that the discriminator incorrectly classifies SR bands. Experimental results obtained on different Sentinel-2 images show the effectiveness of the proposed approach compared to both conventional and deep learning based SR approaches. 
%Since training images are selected randomly over a global range and not limited to specific region, the well-trained S2GAN network can be directly applied to Sentinel-2 images with varying characteristics such as rural areas or cities with a dense population and buildings. 
%
%S2GAN achieves satisfying results even on urban areas which contain plentiful details. 
%
\end{abstract}
%
%\vspace{-0.07in}
\section{Introduction}
%\vspace{-0.07in}
\label{sec:intro}
The new generation of satellite multispectral sensors (e.g., WorldView-3 and Sentinel-2) can acquire images with multiple spectral bands with different spatial resolutions. This is mainly due to the storage and transmission bandwidth restrictions~\cite{lanaras2018super}. Accordingly, one of the most important research topics in remote sensing (RS) is to develop methods for super-resolving the lower-resolution bands and having all image bands at the highest spatial resolution. To this end, several super-resolution (SR) methods are introduced in RS. During the last years deep neural networks, in particular convolutional neural networks (CNNs), are found very effective for SR problems. As an example, in~\cite{liebel2016single} SR of multispectral RS images is applied with convolutional layers by utilizing only lower resolution bands (i.e., single image SR). In~\cite{lei2017super}, residual connections are integrated into the single image SR based architecture to enhance SR performance. In~\cite{lanaras2018super}, a SR approach based on deep residual networks is introduced to further utilize higher resolution bands present in RS images unlike the single image SR approaches. Recent works show that generative adversarial networks (GANs) can significantly increase the performance of image enhancement methods in computer vision~\cite{ledig2017photo, wang2018esrgan}. %ledig2017photo
However, there is only a small number of GAN-based SR studies in RS. As an example, in~\cite{liu2018psgan} a particular GAN (PSGAN) framework is utilized to address RS image pan-sharpening problem. The PSGAN significantly improves the performance of conventional pan-sharpening methods. However, it requires a single band panchromatic image and thus is not directly applicable to SR problems. In~\cite{ma2018super}, SRGAN architecture without the batch normalization layers (TGAN) is trained on computer vision images and fine-tuned with RS images to apply the SR. This approach utilizes only RGB image bands and thus limits to apply SR on high dimensional RS images.  
To address these limitations, we propose a GAN based SR approach (S2GAN) on multispectral multi-resolution RS images. In this paper, we mainly focus on the super-resolution of Sentinel-2 images. The proposed approach aims at increasing the spatial resolution of Sentinel-2 bands with 20m and 60m spatial resolutions to accurately provide the fine spatial details. To this end, the proposed S2GAN exploits the Sentinel-2 bands associated to 10m spatial resolution as a guidance for learning the SR task on lower resolution bands. Experimental results confirm that the S2GAN effectively and accurately provides high resolution bands with significant details from low resolution bands by the adversarial training of generator and discriminator networks. To the best of our knowledge, we present the first study on the application of GANs in the framework of Sentinel-2 image SR problems.
%\vspace{-0.07in}
\section{Proposed Super-Resolution Approach}
%\vspace{-0.07in}
\label{sec:method}
\begin{figure*}[t]
	\centering
	\captionsetup{justification=justified,singlelinecheck=false}
	\includegraphics[width=0.98\linewidth]{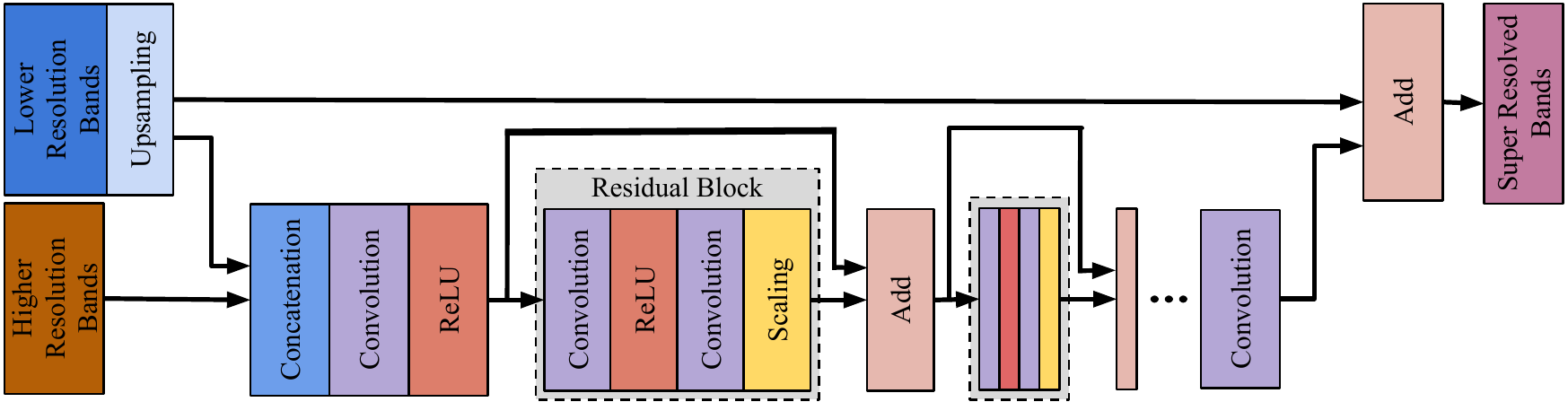}
	\caption{The proposed generator neural network for the characterization of super-resolved bands.}
	\label{fig:G}
\vspace{-0.25cm}
\end{figure*}
Let $I$ be a Sentinel-2 image and $I_{LR}$, $I_{HR}$, $I_{SR}$ be the sets of lower resolution, higher resolution and SR bands, respectively. Sentinel-2 images contain $13$ spectral bands with 10m, 20m and 60m spatial resolutions. Bands 2 to 4 and 8 are associated to 10m resolution, whereas bands 5 to 7, 8A, 11 and 12 have 20m resolution. Remaining bands (1, 9 and 10) are associated to 60m resolution. $I_{HR} $ composes of the spectral bands 2 to 4 and 8, each of which is a section of $W \times H$ pixels. We assume that the set of lower resolution bands can include either bands 5 to 7, 8A, 11, 12 or 1, 5 to 7, 8A, 9 to 12. This can be defined with respect to the scaling factor of the approach, which is either $2$ or $6$. We aim to learn a function $f$, which applies super-resolution on $I_{LR}$ by exploiting $I_{HR}$ as follows:
\begin{equation}
\setlength\abovedisplayskip{7pt}
\setlength\belowdisplayskip{8pt}
\begin{aligned}
 & f: I_{HR}, I_{LR} \to I_{SR} \\
 & \forall \; I_{HR} \in \mathbb{R}^{W\times H\times4} \\
& \forall \; (I_{LR} \in \mathbb{R}^{\frac{W}{2}\!\times\!\frac{H}{2}\!\times6}) \oplus (I_{LR} \in \mathbb{R}^{\frac{W}{2}\!\times\!\frac{H}{2}\!\times6} \times \mathbb{R}^{\frac{W}{6}\!\times\!\frac{H}{6}\!\times3})\\ 
& \exists \; I_{SR} \in \mathbb{R}^{W\times H\times6}
\end{aligned}
\end{equation}
where $I_{SR}$ denotes the SR bands of $I_{LR}$ and $\oplus$ denotes the XOR gate, which results true if only one of the inputs to the gate is true. To this end, we propose a GAN based SR approach, which consists of two main steps: 1) characterization of SR bands by the generator neural network; and 2) classification of SR and ground truth bands by the discriminator neural network. Let $G$ and $D$ be the generator and discriminator networks, respectively. 
$G$ maps the sets of $I_{LR}$ and $I_{HR}$ to the set $I_{SR}$. 
$D$ aims to accurately distinguish generated image bands $I_{SR}$ from their ground truth bands. To this end, we define the adversarial loss over $N$ training images as follows:
\begin{equation}
\setlength\abovedisplayskip{7pt}
\setlength\belowdisplayskip{8pt}
\mathcal{L}_{Adversarial} = \sum_{n=1}^N log(1 - D(G(I_{LR},I_{HR}))).
\end{equation}
$D$ aims to maximize the adversarial loss for better discrimination ability, whereas $G$ aims to minimize this loss to fool discriminator such that discriminator incorrectly labels SR image bands as true bands. 
Thus, this min-max game of $G$ and $D$ is formulated as follows:
\begin{equation}
\setlength\abovedisplayskip{9pt}
\setlength\belowdisplayskip{10pt}
\begin{aligned}
&\!\!\!\!\min_{\theta} \max_{\beta} ~ \mathbb{E}_{I_{GS}\sim p_{data}(I_{GS})} \log D(I_{GS}; {\beta}) + \\
&\!\!\!\mathbb{E}_{(I_{LR},I_{HR})\sim p_G(I_{LR},I_{HR})} \log (1\!-\!D(G(I_{LR},I_{HR}; \theta); \beta))
\label{eq:minmax}
\end{aligned}
\end{equation}
where $\theta$ and $\beta$ are the parameters of generator and discriminator, respectively, and $I_{GS}$ is the set of higher resolution ground truth bands associated to $I_{SR}$. Each step of the proposed approach is explained in the following sections.
%\vspace{-0.15in}
\subsection{Characterization of Super-Resolution Bands}
%\vspace{-0.1in}
This step aims at producing realistic SR image bands, which have similar data distribution as ground truth bands. To obtain the SR image bands, we propose a generator neural network inspired by \cite{lanaras2018super}. 
Different from conventional single image SR approaches, the higher resolution image bands are also utilized in this step together with the lower resolution bands to guide the SR learning approach.   
Thus, the generator learns to transfer information present in higher resolution bands to lower resolution bands. 
To this end, low resolution image bands are first upsampled with the bilinear interpolation and then concatenated with higher resolution bands. 
The subsequent convolution layer, activation layer and 18 Residual blocks are used to extract essential features from combined set of image bands. 
In addition, a long skip-connection between upsampled lower resolution bands and the final output enable the generator network to map the upsampled image bands to the desired higher resolution output. This preserves the radiometry of the input image~\cite{lanaras2018super}. 
In residual blocks, we remove the batch normalization layers. This reduces computational complexity and results in better performance in SR~\cite{wang2018esrgan}. 
The proposed generator neural network is illustrated in Fig.~\ref{fig:G}. It is worth noting that, in addition to the adversarial loss, the pixel-wise mean absolute error (MAE) between the SR and the ground truth bands ($I_{GS}$) is also utilized as the content loss of the generator. 
\vspace{-0.085in}
\subsection{Classification of Super-Resolution Bands}
%\vspace{-0.1in}
\label{sec:D}
\begin{figure*}[t]
	\centering
	\captionsetup{justification=justified,singlelinecheck=false}
	\includegraphics[width=\linewidth]{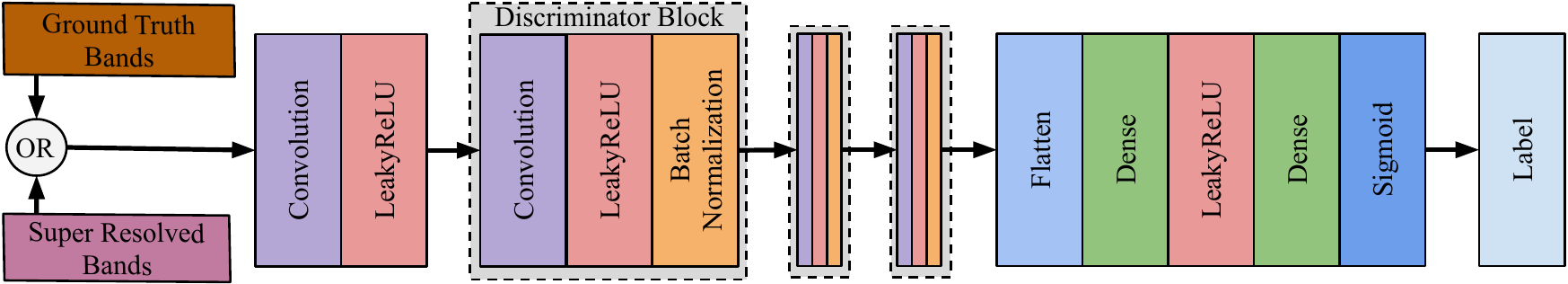}
	\caption{The proposed discriminator neural network for the classification of super-resolved and ground truth bands.}
	\label{fig:D}
\vspace{-0.3cm}
\end{figure*}
This step aims to correctly distinguish SR image bands from their ground truth bands by extracting the high level features for better classification. To this end, this step includes three consecutive blocks, each of which includes a single layer of convolution, activation and batch normalization.
The kernel size of all convolutional layers in the discriminator is $3\times3$. 
Numbers of filters in convolutional layers are 64, 128 and 128. Strides of 2 and 1 are utilized in those layers to reduce the dimensionality of the input. To increase the stability of the adversarial training, Leaky ReLU is used as the activation function of the blocks with batch normalization. Finally, a fully connected layer is included to produce final binary classification probabilities. The proposed discriminator neural network is illustrated in Fig.~\ref{fig:D}. 
\begin{table}[t]
	\setlength{\tabcolsep}{11pt}
	\captionsetup{justification=justified,singlelinecheck=false}
	\caption{SR results (associated to the scaling factor of 2) obtained by the bicubic interpolation, the ATPRK, the SupReME, the Superres, the DSen2 and the proposed S2GAN on the downsampled bands at 20m spatial resolution.}
	\centering	
	\label{tab:full20}
	\begin{tabular}{@{~}lcccc@{~}}
		\toprule
		Method & RMSE & SRE & SAM & UIQ\\
		\midrule
		Bicubic &  123.5 & 25.3 & 1.24 & 0.821 \\
		ATPRK~\cite{wang2015new} &  116.2 & 25.7 & 1.68 & 0.855 \\
		SupReME~\cite{lanaras2017super} &  69.7 & 29.7 & 1.26 & 0.887 \\
		Superres~\cite{brodu2017super} &  66.2 & 30.4 & 1.02 & 0.915 \\
		DSen2~\cite{lanaras2018super}    & 34.5	&36.0 &	0.78	&0.941 \\
		S2GAN & \textbf{33.1} &	\textbf{36.4} &	\textbf{0.74} &	\textbf{0.950}   \\		
		\bottomrule		
	\end{tabular}
\vspace{-0.4cm}
\end{table}
The input to the network is either SR bands from the generator network or the corresponding ground truth bands, whereas the output is the label, which denotes whether the input is ground truth or SR bands ($I_{SR}$). Accordingly, to define the discriminator loss, we incorporate the adversarial loss with the following discriminator loss:
\begin{equation}
\setlength\abovedisplayskip{3pt}
\setlength\belowdisplayskip{4pt}
\mathcal{L}_{Discriminator} = \sum_{n=1}^N log(1-D(I_{GS})).
\end{equation}
If the input data is the ground truth bands of the $I_{SR}$, the output value will be close to 1. In this case, the input has a large probability to be realistic. 
\vspace{-0.07in}
\section{Experimental Results}
%\vspace{-0.07in}
\label{sec:results}
Experiments were conducted on different Sentinel-2 images provided in \cite{lanaras2018super}. We used the same training, validation and test images as suggested in \cite{lanaras2018super}.
%Fig.~\ref{fig:dataset example} shows an example of the images used in the experiments. 
%\begin{figure}[t]
%	\centering	
%	\includegraphics[width=0.15\textwidth]{01.png} \vspace{0.01in}
%	\includegraphics[width=0.15\textwidth]{05.png} \vspace{0.01in}
%	\includegraphics[width=0.15\textwidth]{03.png} \vspace{0.01in}
%	%\includegraphics[width=0.15\textwidth]{04.png}
%	%\includegraphics[width=0.15\textwidth]{02.png}
%	%\includegraphics[width=0.15\textwidth]{06.png}
%	\captionsetup{justification=justified,singlelinecheck=false}
%	\caption{An example of Sentinel-2 images considered in the experiments.}
%	\label{fig:dataset example}
%\vspace{-0.6cm}
%\end{figure}
To optimize the loss functions, we used the mini-batches of size 128 throughout 56 epochs. At each iteration, the generator and the discriminator networks were trained sequentially on NVIDIA Tesla P100 GPU. We compared our approach with: 1) the bicubic interpolation; 2) the area-to-point regression kriging (ATPRK)~\cite{wang2015new} that is a pan-sharpening based approach; 3) the Super-Resolution for Multispectral Multiresolution Estimation (SuperReME)~\cite{lanaras2017super} approach; 4) the Superres~\cite{brodu2017super} that is a geometrical model based approach; and 5) the DSen2~\cite{lanaras2018super} that is a CNN based approach.
Results of each approach are provided in terms of four performance evaluation metrics: 1) Root Mean Squared Error (RMSE), 2) Signal to Reconstruction Error Ratio (SRE), 3) Universal Image Quality Index (UIQ) and 4) Spectral Angle Mapper (SAM). 
SRE measures the error relative to the mean intensity of a SR image band, and thus provides values in decibels (dB). 
UIQ evaluates the luminance, contrast, and structure of a SR image band with the maximum value of 1. 
SAM measures the angular deviation between the spectral signatures of the ground truth and SR bands, and thus provides the values in degrees. 

\begin{figure*}[t]
	\centering
	\captionsetup{justification=justified,singlelinecheck=false}
	\includegraphics[width=0.93\linewidth]{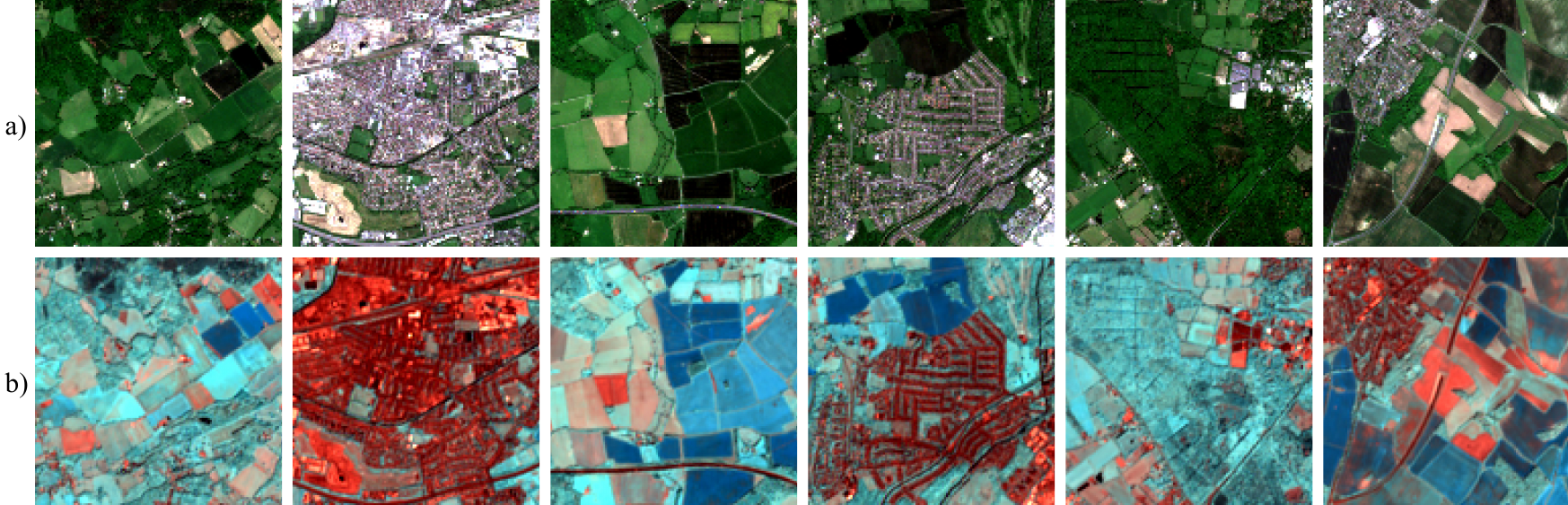}
	\caption{An example of SR results (associated to the scaling factor of 2) obtained by the proposed S2GAN on the downsampled bands associated with 20m spatial resolutions. (a) The true color composite of downsampled RGB bands associated with 10m spatial resolution. (b) The false color composite of SR bands 5 to 7 obtained by the S2GAN.}
	\label{fig:20true}
\vspace{-0.3cm}
\end{figure*}

We applied SR to the bands associated with 20m and 60m spatial resolutions. Due to the unavailability of ground truth bands at 10m spatial resolution for these bands, we followed the downsampling strategy to train the proposed architecture and to evaluate the performance of the S2GAN. To this end, the bands associated with 20m and 60m spatial resolutions were downsampled to 40m and 360m spatial resolutions, and then SR was applied by the considered methods. The average results over all test images associated to the scaling factor of 2 are given in Table~\ref{tab:full20}. As we can see from the table, our approach (S2GAN) performs better than the other approaches under all metrics. These results show that our approach effectively applies SR on the lower resolution Sentinel-2 bands to accurately enhance their spatial resolutions similar to the ground truth bands. To visually evaluate the performance of the S2GAN, we selected test images, which include relatively high subtle details. Fig.~\ref{fig:20true} shows the true color composite of RGB bands and the false color composite of SR bands obtained by the S2GAN. In addition, Table~\ref{tab:city} presents the average RMSE values for SR bands over these test images obtained by the DSen2 and the S2GAN. In such a relatively difficult scenario, the performance of the S2GAN for SR task is more significant compared to the DSen2. This also shows the success of our approach over the state-of-the-art approaches. 

\vspace{-0.1in}
\section{Conclusion}
%\vspace{-0.08in}
\label{sec: conclusion}
This paper proposes a GAN based approach (S2GAN) to enhance the spatial resolution of multispectral multi-resolution Sentinel-2 images. The proposed approach consists of two main steps: 1) accurately increasing the spatial resolution of the bands at 20m and 60m spatial resolutions with the guidance of the bands associated to 10m spatial resolution by the generator neural network; and 2) effectively distinguishing the SR bands from their ground truth bands by the discriminator neural network. We also applied the adverserial learning of generator and discriminator networks.
Experimental results obtained on the Sentinel-2 images indicate that our approach achieves promising performance for the SR of Sentinel-2 lower resolution bands with respect to the state-of-the-art approaches. 
We would like to note that the S2GAN approach can be also applied to any other RS image.
As a future work, we plan to improve the network structures of generator and discriminator steps, which can be achieved by integrating the realistic discriminator or Wasserstein GAN into our approach.
\begin{table}[t]
	\setlength{\tabcolsep}{5.5pt}	
	%\captionsetup{justification=justified,singlelinecheck=false}
	\caption{RMSE (associated to the scaling factor of 2) obtained by the DSen2 and the proposed S2GAN on the downsampled bands at 20m spatial resolution of test images given in Fig.~\ref{fig:20true}.}
	\centering
	\label{tab:city}
	\begin{tabular}{@{~}lccccccc@{~}}
		\toprule
		 Method &  B5 &B6 &B7&B8A&B11 &B12 & Avg.\\ \midrule
		DSen2 &25.3   &51.0 &		61.2&		63.2&		33.4&		30.9	& 44.2\\	
		S2GAN  &\textbf{23.1}   &\textbf{43.4}  &\textbf{52.4}  &\textbf{53.1}  &\textbf{30.7}	&\textbf{28.3} &\textbf{38.5}\\
		\bottomrule
	\end{tabular}
\vspace{-0.4cm}
\end{table}
\vspace{-0.1in}
\section{Acknowledgments} %\footnotesize 
%\vspace{-0.01in}
This work was supported by the European Research Council under the ERC Starting Grant BigEarth-759764. The authors would like to thank Yakun Li, DFKI GmbH, Germany and Dr. Hua Yang​, Shanghai Jiao Tong University, China for the helpful suggestions. 
\vspace{-0.07in}
\bibliographystyle{IEEEbib}
\bibliography{defs,Paper}
\end{document}